\documentclass[preprint superscriptaddress, twocolumn, showpacs, a4paper, UTF8]{revtex4}
\usepackage{bbm}
\usepackage[utf8]{inputenc}
\usepackage[T1]{fontenc}
\usepackage[colorlinks=true,linkcolor=black,citecolor=blue,urlcolor=blue]{hyperref}
\usepackage{longtable}
\usepackage{multirow}
\usepackage{graphicx}    
\usepackage{bm}          
\usepackage{dcolumn}    
\usepackage{amssymb}
\usepackage{color}
\usepackage{CJK}
\usepackage{url}
\usepackage{booktabs}
\usepackage[online,flushleft]{threeparttable}
\usepackage{amsmath,amssymb,mathrsfs,amsfonts,nicefrac}
\usepackage{mathtools}
\usepackage{xcolor}

\begin{document}

\title{Investigation of Langdon effect on the stimulated backward Raman and Brillouin scattering}
\def\athsep{and}
\author{Jie Qiu}
\affiliation{Institute of Applied Physics and Computational
Mathematics, Beijing, 100094, China}
\author{Liang Hao \footnote{Corresponding author: hao\_liang@iapcm.ac.cn}}
\affiliation{Institute of Applied Physics and Computational
Mathematics, Beijing, 100094, China}
\author{Li Hua Cao}
\affiliation{Institute of Applied Physics and Computational
Mathematics, Beijing, 100094, China}
\author{Shiyang Zou}
\affiliation{Institute of Applied Physics and Computational
Mathematics, Beijing, 100094, China}

\date{\today}

\begin{abstract}
In a laser-irradiated plasma, the Langdon effect makes the electron
energy distribution function (EEDF) tend to a super-Gaussian
distribution, which has important influences on
laser plasma instabilities. 
In this work, the influences of a super-Gaussian EEDF on the
convective stimulated backward Raman scattering (SRS) and stimulated
backward Brillouin scattering (SBS) are studied systematically for a
wide range of typical plasma parameters in the inertial confinement
fusion (ICF). Distinct behaviors are found for SRS and SBS in the
variation trend of the peak spatial growth rate and the
corresponding wavelength of the scattered light.
Especially, the Langdon effect on the SBS in plasmas with different
ion species and isotopes is analyzed in detail, and the parameter
boundary for judging the variation trend of the peak spatial growth
rate of SBS with the super-Gaussian exponent is presented for the
first time. In certain plasma parameter region, it is found that the
Langdon effect could enhance SBS in mixed plasma, which may
attenuate the improvement in suppressing SBS by mixing low-Z ions
into the high-Z plasma. These comprehension of Langdon effect on
LPIs would contribute to a better understanding of SRS and SBS in
experiments.
\end{abstract}


\maketitle


\section{Introduction}
In laser-driven inertial confinement fusion (ICF), the dominant
heating mechanism is inverse bremsstrahlung heating. When the
heating rate of inverse bremsstrahlung exceeds the electron
thermalization rate, the electron energy distribution function
(EEDF) would deviate from a Maxwellian EEDF and become a
super-Gaussian one~\cite{Langdon1980BremssEDF,Matte1988NonMax}. This
so-called Langdon effect is tightly related to laser intensity
$I_0$, electron temperature $T_{\rm e}$ and charge state $Z$ of
plasma, and has important influences on the process of electron
conduction, inverse bremsstrahlung absorption, laser plasma
instabilities. Recent study showed that with the increasing
intensity, the Langdon effect occurred when $Zv_{\rm os}^2/v_{\rm
the}^2\gg 1$ but would be suppressed by the nonlinear effect of high
laser intensity when $v_{\rm os}\gg v_{\rm the}$, where $v_{\rm os}$
and $v_{\rm the}$ are the electron quiver velocity and electron
thermal velocity respectively~\cite{Weng2009InvBremNonMax}. In ICF,
the Langdon effect should be important in high-Z plasma ablated from
the hohlraum wall in indirect-drive experiments. Besides, recent
experimental results diagnosed the presence of super-Gaussian EEDF
when multibeams irradiate at the low-Z gas plasma
target~\cite{Turnbull2020LangdonEffectCBET} via its influence on the
spectrum of Thomson
scattering~\cite{Zheng1997ThomsonNMax,Milder2020EEDFInvBHeat}. The
power transfer of the crossed beams was found to be overestimated by
the standard Maxwellian calculations, and the impact of Langdon
effect on the crossed beam energy transfer (CBET) was considered to
be a possible reason for the discrepancy between the Maxwellian
calculations and experimental results on NIF
facility~\cite{Turnbull2020LangdonEffectCBET}.

Besides CBET, backward stimulated Raman scattering (SRS) and
backward simulated Brillouin scattering (SBS) are also important
laser plasma instabilities (LPIs) in ICF. Commonly, SRS and SBS are
three-wave interaction processes where an incident electromagnetic
wave (EMW) decays into a backscattered EMW and a forward propagating
electron plasma wave (EPW) or ion acoustic wave (IAW), respectively.
They are detrimental to inertial confinement fusion, since the
backscattered light can take energy away from the incident laser,
and the hot electrons generated by SRS can preheat the
capsule~\cite{Lindl2004ICFIndirectIgn}. Besides, the strong
backscattered light of SBS also has a potential risk to damage the
optical device within the laser facility. In experiments, the
spectra and reflectivity of the backscattered light are important
diagnostics for SRS and SBS. Currently, the investigation of the
scattered spectra and reflectivity usually assumes a Maxwellian
EEDF~\cite{Hao2014SRSSBSScatter,Strozzi2008RayBackScatter}. However,
there was some discrepancy between the experimentally measured SRS
spectra and the simulated result by the ray-tracing
method~\cite{Hall2017GasFillNIF,Strozzi2017InterPlayLPIHydro}. To
match the experimental reflectivity of SRS, an artificial seed of
SRS backscattered light was also needed in
calculations~\cite{Strozzi2017InterPlayLPIHydro}. Because the SRS
and SBS are very sensitive to the EEDF, one possible source of the
discrepancy is the assumption of the Maxwellian EEDF in current
physics model of the ray-tracing method.
Although the impact of super-Gaussian EEDF on the threshold of SRS
and the ion acoustic frequency of SBS were studied for some specific
plasma conditions
\cite{Bychenkov1997SRSNonMax,Afeyan1998KinNonuniformHeating}, for
practical interest, it is necessary to study the effects of
super-Gaussian EEDF on the gain and spectra of SRS and SBS
systematically for a wide range of typical plasma conditions and
plasma compositions in ICF.

To help resolve these issues, in this work, the influences of
super-Gaussian EEDF with different exponents on the SRS and SBS for
a wide range of plasma conditions are studied systematically based
on the linear gain coefficient calculation. Besides the distinct
behaviors in wavelength of scattered light and the growth rate of
SRS and SBS, some of which are consistent with previous works, the
effects of super-Gaussian EEDF on the peak value of spatial grow
rate of SBS are found to be different in plasma with different ion
compositions. For each typical plasma composition, the boundary of
plasma parameters for the variation trend of the peak spatial growth
rate of SBS with the super-Gaussian exponent is presented for the
first time, which is useful for the convenient judgement of the
variation trend of SBS. One interesting result is that the Langdon
effect can weaken SBS in single species plasma but enhance SBS in
mixed plasma under certain plasma parameter conditions. For such
parameters, although mixing low-Z ion species is often used to
suppress
SBS~\cite{Neumayer2008SupSBSMultIon,Neumayer2008MultipleIon},
Langdon effect may attenuate this improvement. Furthermore, it is
also found that the isotopic type of the low-Z mixture can have a
significant impact on SBS. This work not only advances the further
physical modeling and ray-tracing simulations of backscattering
instabilities with the consideration of Langdon effect, but also
helps to understand the spectra and growth of SRS and SBS in
experiments better.

This paper is organized as follows: In Section~\ref{sec:anay}, the
theoretical analysis method is specified. In
Section~\ref{sec:Result}, the influence of a super-Gaussian EEDF on
the linear gain coefficient versus wavelength is discussed for SRS
and SBS processes under various plasma conditions. In
Section~\ref{sec:conc}, the conclusions as well as some discussions
are given.

\section{Theoretical analysis method}
\label{sec:anay} Due to the inverse Bremsstrahlung (IB) Langdon
heating effect in a laser-irradiated plasma, the slow-variation
component of the electron energy distribution function (EEDF) can
deviate from the Maxwellian~\cite{Langdon1980BremssEDF}. This
distorted EEDF can be described by a super-Gaussian
function~\cite{Matte1988NonMax},
\begin{equation}
  f_{\rm e0}(v)=\frac{n_{\rm e}m}{4\pi v_{\rm the}^3\beta_{m}^3\Gamma(3/m)}\exp[-(\frac{v}{\beta_m v_{\textrm{the}}})^{m}],
  \label{eq:fGauss}
\end{equation}
where the distortion is characterized by the super-Gaussian exponent
$m$, the factor $\beta_{m}=\sqrt{3\Gamma(3/m)/\Gamma(5/m)}$,
$\Gamma$ is the gamma function, and $v_{\rm the}=\sqrt{T_{\rm
e}/m_{\rm e}}$ is the electron thermal velocity. The super-Gaussian
exponent $m$ is determined by the competition between Langdon
heating and electron-electron collisions. By fitting to the
Fokker-Planck simulation, a formula of $m$ which is valid
from low to very high intensity is given as
~\cite{Weng2009InvBremNonMax}
\begin{equation}
  m=2+\frac{3}{1+0.62/\alpha_{\rm Ln}},
  \label{eq:mGauss2}
\end{equation}
where $\alpha_{\rm Ln}=Z_{\rm eff}(v_{\rm os}^2/v_{\rm
the}^2)/(v_{\rm os}^2/v_{\rm the}^2+1)^2$, $Z_{\rm
eff}=\overline{Z^2}/\overline{Z}$, and $v_{\rm os}\equiv eE_0/m_{\rm
e}\omega_0$ is the electron quiver velocity with $E_0$ and
$\omega_0$ as the amplitude and frequency of the incident laser,
respectively.

As resonant instabilities, SRS and SBS are sensitive to the EEDF and
background plasma parameters. To investigate the impact of
super-Gaussian EEDF on SRS and SBS with different $m$ and plasma
parameters, we start from the linear gain exponent which is an
important quantity to characterize the convective amplification of
SRS and SBS~\cite{Lindl2004ICFIndirectIgn} and its spectrum had been
widely used to calculate the backscattered SRS and SBS spectra via
the ray-tracing method~\cite{Strozzi2008RayBackScatter}.
In WKB approximation,
the spectrum of linear gain exponent is defined as
\begin{equation}
  \mathrm{G}_{\rm R,B}(\omega_{\rm s})=\int_{\rm path}{\rm K}_{\rm R,B}(z,\omega_{\rm s}) dz,
\label{eq:kinGain}
\end{equation}
where the subscript $\rm R$ or $\rm B$ denotes the SRS or SBS process, and $\omega_s$ is the radian frequency of the scattered light.
The integration is along one specified laser ray,
and the local spatial growth rate ${\rm K}_{\rm R,B}$ obtained by the kinetic theory~\cite{Drake1974ParaInstabEM} is given by,
\begin{equation}
  {\rm K}_{\rm R,B}(\omega_{\rm s})=\frac{1}{4}\frac{k_{\mathrm{l},a}^2v_{\rm os}^2}{k_{\rm s}c^2}\mathrm{Im}\left[\frac{\chi_{\rm e}(1+\chi_{\rm ion})}{\epsilon}\right]
  \label{eq:kappaz}
\end{equation}
where the subscript $\rm l$ or $a$ is for EPW or IAW respectively,
$\chi_{\rm e}(\omega_{\mathrm{l},a},k_{\mathrm{l},a})$ is the electron susceptibility,
$\chi_{\rm ion}(\omega_{\mathrm{l},a},k_{\mathrm{l},a})$ is the
total susceptibility of ions, and
$\epsilon=1+\chi_{\rm e}+\chi_{\rm ion}$ is the dielectric
function. The frequency $\omega_{\rm s}$ and wave-number $k_{\rm s}$
for the scattered wave are related to $\omega_{\mathrm{l},a}$ and
$k_{\mathrm{l},a}$ for EPW or IAW by the frequency and wave-number
matching condition
\begin{equation}
\begin{aligned}
   \omega_0&=\omega_{\rm s}+\omega_{\rm l,a} \\
   {k}_0&={k}_{\rm s}+{k}_{\rm l,a}
\end{aligned}
\label{eq:wkmatch}
\end{equation}
where $k_0$ is the wavenumber of the incident laser. The incident
laser and scattered light satisfy the dispersion relation for the
EMWs,
\begin{equation}
  \omega_{\rm 0,s}^2=\omega_{\rm pe}^2+c^2 k_{\rm 0,s}^2.
  \label{eq:dispMW}
\end{equation}

The linear kinetic electron susceptibility is obtained from the
perturbed Vlasov equation~\cite{Chen1984Plasma}
\begin{equation}
  \begin{aligned}
    \frac{\partial \delta f_{\rm e}}{\partial t}+\mathbf{v}\cdot \nabla \delta f_{\rm e}& =\frac{e\mathbf{E}}{m}\cdot\frac{\partial f_{\rm e}}{\partial \mathbf{v}} \\
    \delta n_{\rm e} = \int \delta f_{\rm e} d\mathbf{v}&=(\chi_{\rm e}/e)\epsilon_0\nabla\cdot\mathbf{E}
  \end{aligned}
  \label{eq:kinSusE}
\end{equation}
where $f_{\rm e}$ is the EEDF of the background plasma, $\delta
f_{\rm e}$ and $\delta n_{\rm e}$ are the perturbation of EEDF and
electron density respectively, $\mathbf{E}$ is the electrostatic field, $e$ is the charge of electron, and $\epsilon_0$ is the vacuum
permittivity. Notice that the background EEDF includes both a
slow-variation component $f_{\rm e0}$ and the high frequency
components such as $f_{\rm e1}$ which oscillates at $\omega_0$.
However, only $f_{\rm e0}$ can participate in the generation of EPW
or IAW with both the perturbation $\delta n_{\rm e}$ and
$\mathbf{E}$ at frequency $\omega_{\rm l,a}$, since the Gauss's law
$\epsilon_0\nabla\cdot \mathbf{E}=-e\delta n_{\rm e}+\sum
Z_{i\alpha}\delta n_{\rm i\alpha}$, which requires resonance of
$\mathbf{E}$ and $\delta n_{\rm e}$, can not be satisfied for the
density perturbation at the frequency $\omega_0\pm\omega_{\rm l,a}$
driven by $\mathbf{E}$ and the high frequency component $f_{\rm
e1}$. In practical ray-tracing simulation, the main pulse are
divided into different time slices with interval such as $100$ps
($\sim 10^5$ laser period), and the gain ${\rm G}_{\rm R,B}$
calculated at an instantaneous step within each time slice is used
to evaluate the average level of SRS and SBS in the corresponding
time
slice~\cite{Hao2014SRSSBSScatter,Hall2017GasFillNIF,Strozzi2017InterPlayLPIHydro}.
Consequently, $f_{\rm e}$ can be substituted by $f_{\rm e0}$ which
is assumed to be constant within each time slice in calculating
$\chi_{\rm e}$. And
\begin{eqnarray}
  \chi_{\rm e}(\omega,k)&=&{\frac{\omega_{\rm pe}^{2}}{k^2\int f_{\rm e0}d\mathbf{v}}}\int \frac{\mathbf{k}\cdot{\partial f_{\rm e0}}/{\partial \mathbf{v}}}{\omega-\mathbf{k}\cdot
  \mathbf{v}}\mathrm{d}\mathbf{v} \nonumber\\
  &=&{\frac{\omega_{\rm pe}^{2}}{k^2\int_{-\infty}^{\infty} f_{\rm e0}^xdv_x}}\int_{-\infty}^{\infty} \frac{{\partial f_{\rm e0}^x}/{\partial v_x}}{\omega/k-v_x}dv_x
\label{eq:xe1d}
\end{eqnarray}
where
$\omega_{\rm pe}$ is the electron plasma frequency,
and the 1D distribution function $f_{\rm e0}^x(v_x)=\iint f_{\rm e0}(v_x,v_y,v_z)dv_ydv_z$.
For the isotropic super-Gaussian EEDF given in Eq.~(\ref{eq:fGauss}), it can be obtained
\begin{equation}
  \chi_{\rm e}(\omega,k,m)=\frac{1}{k^2\lambda_{\rm De}^2}\mathcal{Z}_{\rm e}[\frac{\omega}{kv_{\rm the}},m]
  \label{eq:chieGauss}
\end{equation}
where
\begin{equation}
  \mathcal{Z}_{\rm e}[x,m]\equiv \frac{1}{A_m}\left[1+\frac{m}{2\Gamma(1/m)}\frac{x}{\beta_m}\int_{-\infty}^{\infty} \frac{\exp(-u^m)}{u-\frac{x}{\beta_m }}du\right],
  \label{eq:chieGaussFem}
\end{equation}
where $A_m=3\Gamma^2(3/m)/[\Gamma(1/m)\Gamma(5/m)]$.

A Maxwellian distribution with an ion temperature $T_{\rm i}$ is assumed for all the ion species, yielding the ion susceptibility,
\begin{equation}
  \chi_{\rm ion}=\sum_\alpha\chi_{\rm i\alpha}(\omega,k)=\sum_\alpha\frac{1}{k^2\lambda_{\rm D\alpha}^2}[1+\zeta_\alpha \mathcal{Z}(\zeta_\alpha)]
  \label{eq:xion}
\end{equation}
summed over the ion species $\alpha$, where $\mathcal{Z(\zeta_\alpha)}$ is the
plasma dispersion function with $\zeta_\alpha\equiv
{\omega}/{\sqrt{2}v_{\rm th\alpha}k}$, the thermal velocity $v_{\rm
th\alpha}=\sqrt{T_{\rm i}/m_\alpha}$, the Debye length $\lambda_{\rm
D\alpha}=v_{\rm th\alpha}/\omega_{\rm p\alpha}$, and the ion plasma
frequency $\omega_{\rm p\alpha}=\sqrt{{e^2Z_{\alpha}^2n_{\rm
i\alpha}}/{\epsilon_0m_\alpha}}$ with charge state $Z_{\alpha}$ and
ion mass $m_{\alpha}$.

Since the linear gain exponent and the time-resolved spectra of
backscattered light of SRS and SBS can be obtained by integrating
the spatial growth rate ${\rm K}_{\rm R,B}$ along one ray path with
specified practical profiles of plasma parameters usually provided
by the radiation hydrodynamic simulations, as done in the
ray-tracing
method~\cite{Hao2014SRSSBSScatter,Strozzi2008RayBackScatter}, we
mainly analyze the impact of $m$ on ${\rm K}_{\rm R,B}$ given in
Eq.~(\ref{eq:kappaz}) to identify the behaviors under different
plasma parameters and plasma composition in following section.

\section{Influences of Langdon effect on SRS and SBS}
\label{sec:Result} 
\begin{figure}[!h]
  \centering
  \includegraphics[angle=0,width=0.4\textwidth]{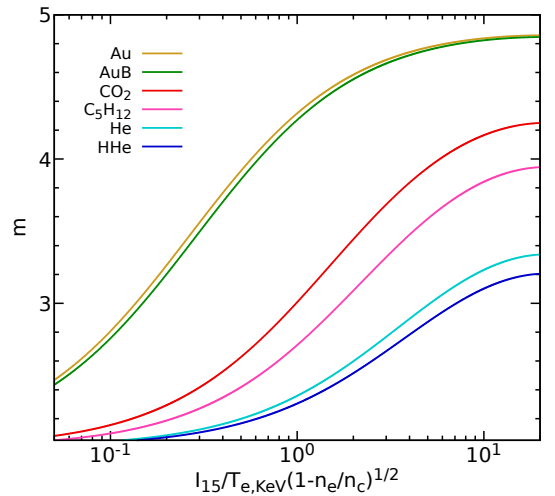}
  \caption{
     The variation of $m$ with $I_{15}/T_{\rm e,KeV}(1-n_{\rm e}/n_{\rm c})^{1/2}$ for He ($Z=2$), HHe ($Z_{\rm eff}\approx 1.67$),
     C$_5$H$_{12}$ ($Z_{\rm eff}\approx 4.57$),
     CO$_{2}$ ($Z_{\rm eff}\approx 7.45$),
     Au ($Z=50$), and AuB ($Z_{\rm Au}=50$, $Z_{\rm B}=5$, $Z_{\rm eff}\approx 45.9$) plasma
     in practical parameter range of hohlraum experiment.
  }
  \label{Fig:Gaussm0}
\end{figure}

In ICF hohlraum, the evolution of background
plasma is determined by all the irradiating beams. The typical ion
species includes the initial filled gas like mid-Z $\rm C_5H_{12}$
or $\rm CO_2$, or low-Z He or the mixture of $\rm H_2$ and He
(labeled as HHe in following paper), the mid-Z $\rm CH$ plasma from
the ablation off the capsule, and the high-Z plasma such as Au or
AuB ablated from hohlraum
wall~\cite{Lindl2004ICFIndirectIgn,Neumayer2008SupSBSMultIon,Neumayer2008MultipleIon}.
According to the empirical formula given by Eq.~(\ref{eq:mGauss2}),
the super-Gaussian exponent $m$ depends on $Z_{\rm eff}$ and $v_{\rm
os}^2/v_{\rm the}^2\propto I_{15}/T_{\rm e,KeV}(1-n_{\rm e}/n_{\rm
c})^{1/2}$, where $I_{15}=I_0[\rm W/cm^2]/10^{15}$,
$T_{\rm e,KeV}$ is the value of $T_{\rm e}$ in unit of KeV, and
$n_{\rm c}$ is the critical density of incident laser with
wavelength $\lambda_0 = 0.351~\rm \mu m$ in this paper. In hohlraum
experiment, the typical $T_{\rm e}$ is about $\sim \rm KeV$.
Considering the complex beam
overlapping~\cite{Kirkwood2013LPIReview} and the high intensity
speckles in the laser spot~\cite{Lindl2004ICFIndirectIgn}, the local
intensity for IB heating can range from $10^{14}$ to about
$10^{16}~\rm W/cm^2$.
Consequently, the possible values of $I_{15}/T_{\rm e,KeV}(1-n_{\rm
e}/n_{\rm c})^{1/2}$ approximately cover the range of
$0.1\text{-}10$ in ICF hohlraum, over which the variation of $m$
with $I_{15}/T_{\rm e,KeV}(1-n_{\rm e}/n_{\rm c})^{1/2}$ is shown in
Fig.~\ref{Fig:Gaussm0} for the typical ion compositions in hohlraum.
It can be seen that the typical value of the super-Gaussian exponent
$m$ can vary from $2$ to about $3$, $4$ and $5$ for the low-Z, mid-Z
and high-Z plasma, respectively. Consequently, although the plasma
parameters vary at different locations on different rays in
hohlraum, the Langdon effect should be prevalent and important to
LPIs.

%


\subsection{SRS process}
\begin{figure}[!h]
  \centering
  \includegraphics[angle=0,width=0.48\textwidth]{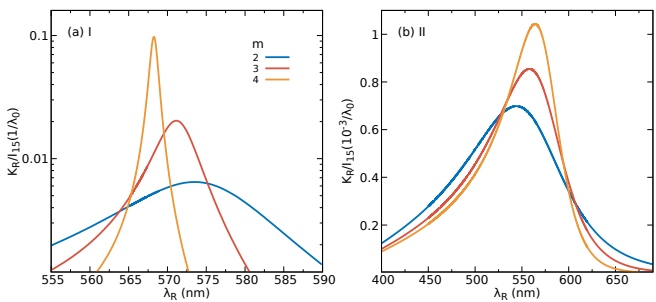}
  \caption{
    The spatial growth rate (${\rm K}_{\rm R}/I_{15}$) versus scattered wavelength ($\lambda_{\rm R}$) at different $m$ for (a) case I and (b) case II.
    In case I, $n_{\rm e}=0.1~n_{\rm c}$, $T_{\rm e}=3~\rm KeV$. In
    case II, $n_{\rm e}=0.05~n_{\rm c}$, $T_{\rm e}=5~\rm KeV$.
  }
  \label{Fig:SRSGain}
\end{figure}

In hohlraum, SRS mainly occurs in the gas region with low-Z or mid-Z
plasma~\cite{Lindl2004ICFIndirectIgn}, for which $m$ can vary from
$2$ to about $4$ as shown in Fig.~\ref{Fig:Gaussm0}. Through a
systematic analysis of the influence of $m$ on $\mathrm{K}_{\rm R}$
for various plasma conditions, it is found that the wavelength of
SRS backscattered light corresponding to the peak value of ${\rm
K}_{\rm R}$ decreases with $m$ in regime of high $n_{\rm e}$ and low
$T_{\rm e}$ while increases with $m$ in regime of low $n_{\rm e}$
and high $T_{\rm e}$, but the peak value of ${\rm K}_{\rm R}$ always
increases with $m$, irrespective of the specific plasma condition.
Besides, the half width of ${\rm K}_{\rm R}$ versus $\lambda_{\rm
R}$ is anti-correlated to the peak value of ${\rm K}_{\rm R}$,
implying that it always decreases with $m$. To illustrate these
effects, as two typical examples in different regimes of $n_{\rm e}$
and $T_{\rm e}$, case I and case II are shown in
Fig.~\ref{Fig:SRSGain}(a) and Fig.~\ref{Fig:SRSGain}(b),
respectively. Notice that $\mathrm{K}_{\rm R}$ is proportional to
the intensity of single beam which stimulates the SRS, while the
single beam intensity may not be the total intensity leading to the
IB heating, due to effects such as beam overlapping. For the
universality, $\mathrm{K}_{\rm R}/I_{15}$ which is independent of
the single beam intensity is used to characterize the physics trends
in these figures.

\begin{figure}[!h]
  \centering
  \includegraphics[angle=0,width=0.48\textwidth]{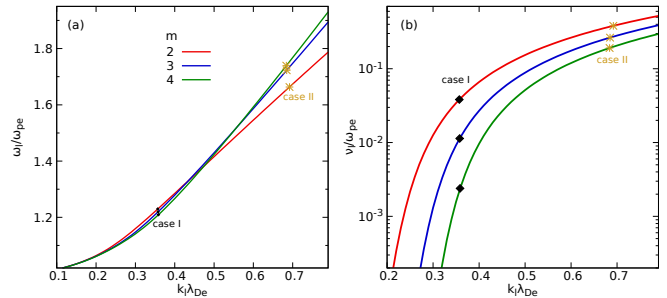}
  \caption{
    (a) The dispersion relations and (b) Landau damping
    of EPW at different $m$. The matching points of case I and case II are indicated by the points for each $m$.
   }
  \label{Fig:SRSDispDamp}
\end{figure}

Commonly, ${\mathrm{K}}_{\rm R}$ peaks approximately at the naturally resonant
frequency of EPW, which satisfies the dispersion equation $\chi_{\rm e}(\omega_{\rm
l}-\mathbbm{i}\nu_{\rm l},k_{\rm l},m)+1=0$, where $k_{\rm l}$,
$\omega_{\rm l}$ and $\nu_{\rm l}$ are the wavenumber, frequency and
Landau damping of EPW, respectively. And the wavelength of SRS
scattered light is related to $\omega_{\rm l}$ by $\lambda_{\rm
R}=\omega_0\lambda_0/(\omega_0-\omega_{\rm l})$. To understand the reason
for the different variation trends of peak wavelength with $m$ at
different plasma conditions, a normalized dispersion relation
\begin{equation}
\mathcal{Z}_{\rm e}[\frac{\omega_{\rm l}-\mathbbm{i} \nu_{\rm l}}{\omega_{\rm
pe}(k_{\rm l}\lambda_{\rm De})},m]+k_{\rm l}^2\lambda_{\rm De}^2=0
\label{eq:LWdis}
\end{equation}
of $\omega_{\rm l}/\omega_{\rm pe}$ versus $k_{\rm l}\lambda_{\rm De}$ for EPW is calculated at different $m$ as shown in
Fig.~\ref{Fig:SRSDispDamp}. For a given $k_{\rm l}\lambda_{\rm De}$, there
are roughly three regimes in Fig.~\ref{Fig:SRSDispDamp}(a). In
regime I, where $k_{\rm l}\lambda_{\rm De}\lesssim 0.43$,
$\omega_{\rm l}$ decreases monotonously with $m$, and the dependence
of the dispersion relation on $m$ is rather weak at about $k_{\rm
l}\lambda_{\rm De}< 0.2$. In regime II, where $0.43< k_{\rm
l}\lambda_{\rm De}\lesssim 0.55$, the variation trend of $\omega_{\rm l}$ with $m$ becomes non-monotonous.
In regime III,
where $k_{\rm l}\lambda_{\rm De}> 0.55$, $\omega_{\rm l}$ increases
monotonously with $m$. In fact, for a fixed $k_{\rm l}\lambda_{\rm De}$,
$\Delta \omega_{\rm l}/\Delta m=-(\partial \mathcal{Z}_{\rm e}/\partial
m)/(\partial \mathcal{Z}_{\rm e}/\partial \omega_{\rm l})$ can be obtained from
Eq.~(\ref{eq:LWdis}). It can be proven $\partial
\mathcal{Z}_{\rm e}/\partial \omega_{\rm l}>0$ at the EPW frequency, but the
sign of $\partial \mathcal{Z}_{\rm e}/\partial m$ depends on the value of
$k_{\rm l}\lambda_{\rm De}$ and $m$, which results in the three regimes of
different $k_{\rm l}\lambda_{\rm De}$. For the case I and case II shown in
Fig.~\ref{Fig:SRSGain}, by solving the dispersion relations
Eq.~(\ref{eq:dispMW}) and Eq.~(\ref{eq:LWdis}) together with the
matching conditions Eq.~(\ref{eq:wkmatch}), $k_{\rm l}\lambda_{\rm De}$
can be obtained to be about $0.36$ for case I and $0.68$ for case II
as indicated in Fig.~\ref{Fig:SRSDispDamp}(a). As expected, case I
and case II are located in the regimes where $\lambda_{\rm R}$ is blue-shifted and red-shifted with $m$, respectively.

\begin{figure}[!h]
  \centering
  \includegraphics[angle=0,width=0.48\textwidth]{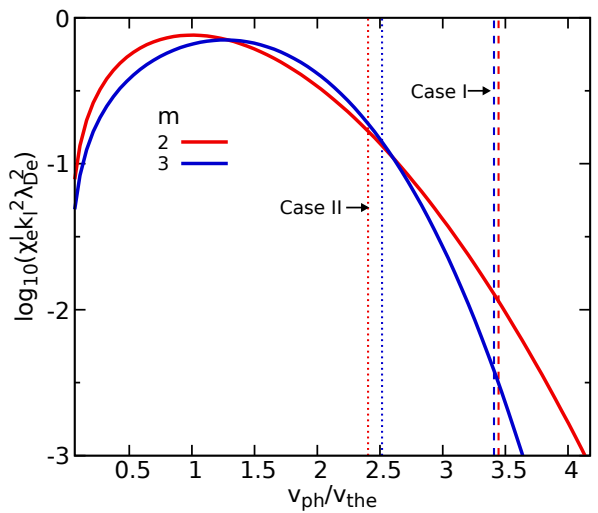}
  \caption{
    $\chi_{\rm e}^{\rm I}k_{\rm l}^2\lambda_{\rm De}^2\propto -\partial f_{\rm e0}^x/\partial v_x$ (see Eq.~\ref{eq:ImXe}) at
    different $m$. The matched EPW phase velocities at each $m$ for case I and case II are indicated by the vertical lines (each $m$ with one color).
  }
  \label{Fig:ffSRS}
\end{figure}

To understand the reason for the behavior of the peak value of
$\mathrm{K}_{\rm R}$ with $m$ and the anti-correlation between the peak
value and the half width of ${\mathrm{K}}_{\rm R}$,
\begin{equation}
  \mathrm{K}_{\rm R}\propto -\mathrm{Im}[\frac{1}{1+\chi_{\rm e}}]=\frac{\chi_{\rm e}^{\rm I}}{(1+\chi_{\rm e}^{\rm R})^2+(\chi_{\rm e}^{\rm I})^2}
  \label{eq:KRchie}
\end{equation}
can be deduced from Eq.~(\ref{eq:kappaz}), where the superscripts
$\rm R$ and $\rm I$ denote the real and imaginary parts,
respectively.
According to Eq.~(\ref{eq:KRchie}), the half width is determined by
$|1+\chi_{\rm e}^{\rm R}|\approx \chi_{\rm e}^{\rm I}$, so a higher
peak value of ${\rm K}_{\rm R}$ is always related to a smaller half
width of it. At the peak, $\chi_{\rm e}^{\rm R}+1\approx 0$, and the peak value
of ${\mathrm{K}}_{\rm R}$ is proportional to $1/\chi_{\rm e}^{\rm I}$. Since
$\chi_{\rm e}^{\rm I}=\nu_{\rm l}(\partial \epsilon^{\rm R}/\partial \omega)\approx
\nu_{\rm l}(\partial \chi_{\rm e}^{\rm R}/\partial
\omega)$~\cite{Strozzi2008RayBackScatter}, it is $\nu_{\rm l}$ that primarily
determines the trend of the peak value of ${\mathrm{K}}_{\rm R}$ with $m$.
As shown in Fig.~\ref{Fig:SRSDispDamp}(b), $\nu_{\rm l}$ decreases with
$m$ for all $k_{\rm l}\lambda_{\rm De}$, so the peak value of ${\mathrm{K}}_{\rm R}$ always
increase with $m$.
The trend of Landau damping with $m$ and $k_{\rm l}\lambda_{\rm De}$ can
be understood from the feature of the EEDF near the matched phase
velocity $v_{\rm ph}=\omega_{\rm l}/k_{\rm l}$. For $\nu_{\rm l} \ll
\omega_{\rm l}$, $\chi_{\rm e}^{\rm I}$ from Eq.~(\ref{eq:xe1d}) can be written as
\begin{eqnarray}\label{eq:ImXe}
\chi_{\rm e}^{\rm I}\approx {\frac{\omega_{\rm
pe}^{2}}{k_{\rm l}^2\int_{-\infty}^{\infty} f_{\rm
e0}^xdv_x}}\int_{-\infty}^{\infty}\frac{\partial f_{\rm
e0}^x}{\partial
v_x}\mathrm{Im}[\frac{1}{\omega_{\rm l}/k_{\rm l}-v_x}]dv_x\\\nonumber
=\frac{-\pi \omega_{\rm pe}^{2}}{k_{\rm l}^2\int_{-\infty}^{\infty} f_{\rm
e0}^xdv_x}\frac{\partial f_{\rm e0}^x}{\partial v_x}|_{v_x=v_{\rm
ph}},
\end{eqnarray}
where second equality follows by using
$\mathrm{Im}[1/(\omega/k-v_x)]=-\pi
\delta(\omega/k-v_x)$~\cite{Chen1984Plasma}. As shown in
Fig.~\ref{Fig:ffSRS}, in case I of small $k_{\rm l}\lambda_{\rm De}$, the
matched $v_{\rm ph}$ is located in the tail of the EEDF where
$\chi_{\rm e}^{\rm I}\propto e^{-v_{\rm ph}^m}$, and $\nu_{\rm l}$ decreases with $m$
rapidly. While in case II of large $k_{\rm l}\lambda_{\rm De}$, the
matched $v_{\rm ph}$ is in the bulk region of the EEDF where
$\chi_{\rm e}^{\rm I}(v_{\rm ph})$ at different $m$ is close. For such a case,
it is the increase of the matched phase velocity with $m$ (see
Fig.~\ref{Fig:SRSDispDamp}a) that leads to the decrease of Landau
damping with $m$.

\subsection{SBS process}
\begin{figure}[!h]
  \centering
  \includegraphics[angle=0,width=0.48\textwidth]{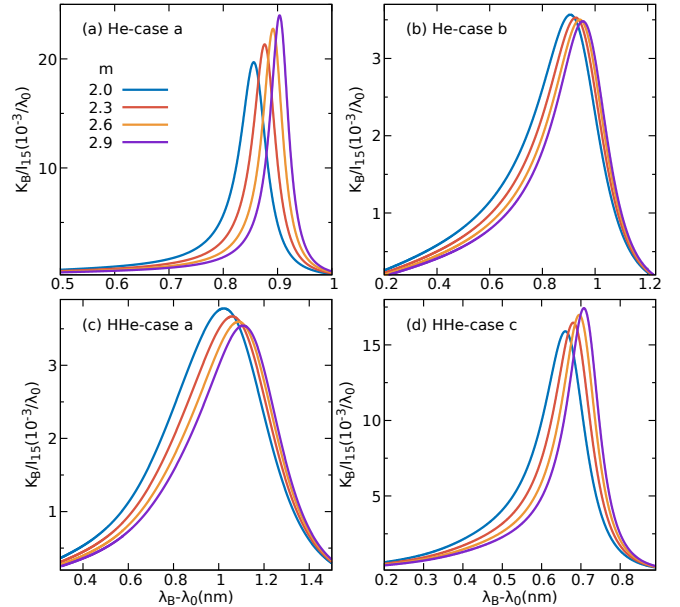}
  \caption{
    The spatial growth rate (${\rm K}_{\rm B}/I_{15}$) of SBS versus the scattered wavelength shift ($\lambda_{\rm B}-\lambda_0$) at different super-Gaussian exponents $m$ for several cases in a fully ionized (a-b) He and (c-d) HHe plasma.
    The case conditions are specified in Table~\ref{Tab:caseHeHHe}.
  }
  \label{Fig:SBSGainHeHHe}
\end{figure}

\begin{figure}[!h]
  \centering
  \includegraphics[angle=0,width=0.48\textwidth]{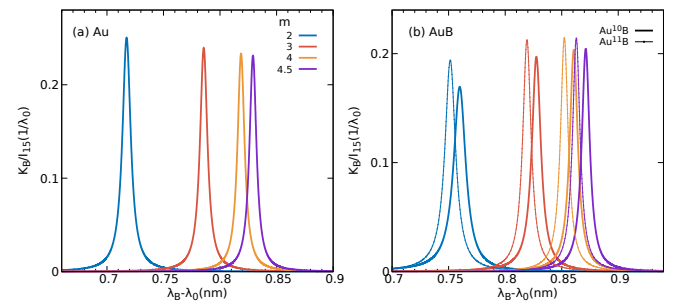}
  \caption{
    The convective spatial growth rate (${\rm K}_{\rm B}/I_{15}$) of SBS versus the scattered wavelength shift ($\lambda_{\rm B}-\lambda_0$) at different super-Gaussian exponents $m$ in (a) Au plasma
    and (b) Au$^{10}$B and Au$^{11}$B plasmas.
    The condition of case `d' as specified in Table~\ref{Tab:caseAuAuB} is taken.
  }
  \label{Fig:SBSGainAuAuB}
\end{figure}

Commonly, SBS can occur in both the gas region with low-Z or mid-Z
plasma and the ablated region with high-Z plasma in
hohlraum~\cite{Lindl2004ICFIndirectIgn}. To illuminate the influence
of super-Gaussian exponent $m$ on SBS, several examples with the
typical parameters in hohlraum are shown in
Figs.~\ref{Fig:SBSGainHeHHe} and~\ref{Fig:SBSGainAuAuB} for the
low-Z and high-Z plasma with the specific conditions listed in
Tables~\ref{Tab:caseHeHHe} and~\ref{Tab:caseAuAuB}, respectively.
In low-Z He and HHe plasma, we take $m=2,2.3,2.6,2.9$ as examples,
while in high-Z Au and AuB plasma, $m=2,3,4,4.5$ are discussed.
The intensity requirements for the corresponding $m$ of super-Gaussian EEDF,
are listed in Tables~\ref{Tab:caseHeHHe} and ~\ref{Tab:caseAuAuB} for these cases,
which are possible due to the high intensity speckles and beam overlapping effects.
Comparing different cases shown in Figs.~\ref{Fig:SBSGainHeHHe}
and~\ref{Fig:SBSGainAuAuB}, it is found that the peak wavelength of
SBS backscattered light always increases with $m$ for a given
parameter conditions and ion composition, but the influence of $m$
on the peak value of ${\rm K}_{\rm B}$ is much more complicated. In
both He and HHe plasmas, the peak value of ${\rm K}_{\rm B}$ can
increase or decrease with $m$ according to the plasma conditions.
Even at the same condition labeled as case `a', the variation trend of
${\rm K}_{\rm B}$ with $m$ can be opposite in He and HHe plasma. In
high-Z case `d', the peak $\mathrm{K}_{\rm B}$ decreases with $m$ in
pure Au plasma, while the trend of the peak $\mathrm{K}_{\rm B}$
with $m$ becomes non-monotonic in AuB plasma, and depends
sensitively on the isotopic type of the low-Z species $\rm B$.
%
\begin{table}[!ht]
\centering
\begin{threeparttable}
\caption{
  The representative cases for He and HHe (1:1) plasma.
  In all cases, the flow velocity is assumed to be zero.
} \label{Tab:caseHeHHe}
\begin{tabular}{ccccccc}
\toprule
\multirow{2}{*}{Case} & \multirow{2}{*}{$n_{\rm e}/n_{\rm c}$} & \multirow{2}{*}{$T_{\rm e}(\rm KeV)$} & \multirow{2}{*}{$T_{\rm i}(\rm KeV)$} & \multicolumn{3}{c}{$I_{15}$\tnote{1}} \\
&  &  &  & $m=2.3$ & $m=2.6$ & $m=2.9$ \\
\midrule
`a'-He & 0.08 & 2.5 & 0.5 & 1.9 & 4.8 & 9.8 \\
`b'-He & 0.06 & 2 & 1 & 1.6 & 3.9 & 7.9 \\
`a'-HHe & 0.08 & 2.5 & 0.5 & 2.4 & 6.0 & 12.9 \\
`c'-HHe & 0.06 & 1 & 0.12 & 0.95 & 2.4 & 5.2 \\
\bottomrule
\end{tabular}
\begin{tablenotes}
\item[1] {
    The required $I_{15}$ for each $m$ is estimated using Eq.~(\ref{eq:mGauss2}).
}
\end{tablenotes}
\end{threeparttable}
\end{table}
\begin{table*}[!ht]
\centering
\begin{threeparttable}
\caption{
  The representative cases for Au and AuB (1:1) plasma.
  In all cases, the flow velocity is assumed to be zero.
} \label{Tab:caseAuAuB}
\begin{tabular}{cccccccc}
\toprule \multirow{2}{*}{Case} & \multirow{2}{*}{$Z_\alpha$} &
\multirow{2}{*}{$n_{\rm e}/n_{\rm c}$} & \multirow{2}{*}{$T_{\rm e}(\rm KeV)$} & \multirow{2}{*}{$T_{\rm i}(\rm KeV)$} & \multicolumn{3}{c}{$I_{15}$\tnote{1}} \\
&  & &  &  & $m=3$ & $m=4$ & $m=4.5$ \\
\midrule
    `d'-Au & $Z_{\rm Au}=50$ & 0.2 & 5 & 0.9 & 0.61 & 2.5 & 6.9 \\
    `d'-AuB & {$Z_{\rm Au}=50$,$Z_{\rm B}=5$} & 0.2 & 5 & 0.9 & 0.67 & 2.8 & 7.6  \\
\bottomrule
\end{tabular}
\begin{tablenotes}
\item[1] {
    The required $I_{15}$ for each $m$ is estimated using Eq.~(\ref{eq:mGauss2}).
}
\end{tablenotes}
\end{threeparttable}
\end{table*}

To comprehend the behavior of ${\rm K}_{\rm B}$ with $m$,
Eq.~(\ref{eq:kappaz}) can be written as
\begin{equation}
  {\rm K}_{\rm B}=-\frac{1}{4}\frac{k_{a}^2v_{\rm os}^2}{k_{\rm s}c^2}\frac{\mathscr{D}^{\rm I}}{(\mathscr{D}^{\rm R})^2+(\mathscr{D}^{\rm
  I})^2},
  \label{eq:KapEstimate}
\end{equation}
where
\begin{equation}
  \mathscr{D}\equiv\frac{1}{\chi_{\rm e}}+\frac{1}{1+\chi_{\rm ion}}
  \label{eq:Ddef}
\end{equation}
and superscripts $\mathrm{R}$ and $\mathrm{I}$ denote the real and
imaginary parts, respectively. Near the peak of ${\rm K}_{\rm B}$,
$\mathscr{D}^{\rm R}\approx 0$ is consistent with the dispersion
relation of IAW $\epsilon\approx 0$ for weak damping $|\chi_{\rm
e}^{\rm I}|\ll |\chi_{\rm e}^{\rm R}|$ and $|\chi_{\rm ion}^{\rm
I}|\ll |1+\chi_{\rm ion}^{\rm R}|$, implying $\mathrm{K}_{\rm B}$
peaks near the natural resonance mode. And the peak value of spatial
growth rate
\begin{equation}
  \mathrm{K}_{\rm B}\approx -\frac{1}{4}\frac{k_{a}^2v_{\rm os}^2}{k_{\rm s}c^2}\frac{1}{\mathscr{D}^{\rm I}}.
  \label{eq:gainpeak}
\end{equation}

To see the reason for the redshift of SBS backscattered light with
$m$, the dispersion relation of IAW is solved.
Under the approximation $\omega_{a}/k_{a}\ll v_{\rm the}$, for the super-Gaussian EEDF
\begin{equation}
  \chi_{\rm e}^{\rm R}\approx \frac{1}{A_mk_{a}^2\lambda_{\rm De}^2} \propto \frac{1}{k_{a}^2}\frac{n_{\rm e}}{A_mT_{\rm e}}.
  \label{eq:chieRLowLimit}
\end{equation}
Because $A_m$ increases with $m$
monotonically, $\chi_{\rm e}^{\rm R}$ decreases with $m$, implying a reduced
number of slow electrons to shield the ions due to the Langdon effect~\cite{Afeyan1998KinNonuniformHeating}.
For such $\chi_{\rm e}^{\rm R}$,
the ion acoustic velocity $c_{\rm s}=\omega_{a}/k_{a}$ in plasma composed of one or
two species~\cite{Williams1995IAWTwoIon,Feng2020Laser3wto2w}, is given by
\begin{equation}
  c_{\rm s}^2=\frac{1}{2}(A\pm \sqrt{B})
  \label{eq:csTwo}
\end{equation}
where ``$+$'' corresponds to the fast mode, and ``$-$'' corresponds
to the slow mode, and
\begin{equation}
  A\equiv (\frac{\gamma_1}{A_1}+\frac{\gamma_2}{A_2})\frac{T_{\rm i}}{M_{\rm p}}+\frac{\overline{Z^2/A}A_mT_{\rm e}}{\overline{Z}M_{\rm p}q}
  \label{eq:defACs}
\end{equation}
and
\begin{eqnarray}\label{eq:defBCs}
&B&\equiv (\frac{\gamma_1}{A_1}-\frac{\gamma_2}{A_2})^2(\frac{T_{\rm i}}{M_{\rm p}})^2
+(\frac{\overline{Z^2/A}A_mT_{\rm e}}{\overline{Z}M_{\rm p}q})^2 \\\nonumber
&+&2(\frac{\gamma_1}{A_1}-\frac{\gamma_2}{A_2})(\frac{f_1Z_1^2}{A_1}-\frac{f_2Z_2^2}{A_2})\frac{T_{\rm
i}A_mT_{\rm e}}{M_{\rm p}^2\overline{Z}q}.
\end{eqnarray}
Here $q\equiv 1+A_mk_{a}^2\lambda_{\rm De}^2$, $M_{\rm p}$ is the
proton mass, the ion species is indicated by the subscript $\alpha=1,2$,
$\gamma_\alpha$ is the adiabatic exponent, $A_\alpha$ is the mass number, and $f_\alpha$ is the number fraction of ion species $\alpha$. The bars denote the average, i.e.,
\begin{align*}
 & \overline{Z}=f_1Z_1+f_2Z_2  \\\nonumber
 & \overline{Z^2/A}=f_1Z_1^2/A_1+f_2Z_2^2/A_2.
  \label{eq:avedef}
\end{align*}
In the expression of $c_{\rm s}$, the term $A_mT_{\rm e}$ appears as a whole,
so it can be thought that
the Debye shielding electron temperature is effectively boosted to
$A_mT_{\rm e}$. In Appendix~\ref{app:csprove}, it is proven that
$c_{\rm s}$ increases with $A_mT_{\rm e}$ for both the fast and slow modes,
when all other parameters such as $T_{\rm i}$ and $n_{\rm e}$ are
kept constant. So it can be expected that due to the effective
increase of Debye shielding electron temperature caused by a
super-Gaussian EEDF, the IAW frequency $\omega_{a}=k_{a} c_{\rm s}$
increases with $m$. Consequently, the scattered wavelength
$\lambda_{\rm B}=\omega_0\lambda_0/(\omega_0-\omega_a)$ increases as
$m$ increases.


From Eq.~(\ref{eq:gainpeak}),
the behavior of the peak value of ${\rm K}_{\rm B}$
with $m$ is primarily determined by the value of $-\mathscr{D}^{\rm I}$ near the peak wavelength.
Since at the peak $\epsilon\approx 0$ and
$|\chi_{\rm e}|\approx |1+\chi_{\rm ion}|$, it is found ${\rm K}_{\rm B}\propto-1/\mathscr{D}^{\rm I}\approx |\chi_{\rm e}||1+\chi_{\rm
ion}|/(\chi_{\rm e}^{\rm I}+\chi_{\rm ion}^{\rm I})$. Considering
that $\chi_{\rm e}^{\rm I}+\chi_{\rm ion}^{\rm
I}=\mathrm{Im}[\epsilon]=(\nu_{\rm e}+\nu_{\rm ion})\partial
\epsilon/\partial \omega$, where $\nu_{\rm e}$ and $\nu_{\rm ion}$
are electron and ion Landau damping rates respectively, and
$|\chi_{\rm e}||1+\chi_{\rm ion}|$ is proportional to the
ponderomotive drive for IAW~\cite{Strozzi2008RayBackScatter}, it can
be understood that ${\rm K}_{\rm B}$ is determined by the tradeoff
between the ponderomotive drive and the Landau damping of IAW. We
can also separate out the ion and electron contributions to
$\mathscr{D}^{\rm I}$,
\begin{equation}
\mathscr{D}^{\rm I}=\mathscr{D}_{\rm ion}^{\rm I}+\mathscr{D}_{\rm e}^{\rm I}
  \label{eq:DIsum}
\end{equation}
with
\begin{equation}
\mathscr{D}_{\rm ion}^{\rm I}\equiv \mathrm{Im}[1/(1+\chi_{\rm ion})]=-\chi_{\rm ion}^{\rm I}/|1+\chi_{\rm ion}|^2
  \label{eq:Dion}
\end{equation}
and
\begin{equation}
  \mathscr{D}_{\rm e}^{\rm I}\equiv \mathrm{Im}[1/\chi_{\rm e}]=-\chi_{\rm e}^{\rm I}/|\chi_{\rm e}|^2.
\label{eq:De}
\end{equation}
At the peak of ${\rm K}_{\rm B}$, $\mathscr{D}_{\rm ion}^{\rm
I}/\mathscr{D}_{\rm e}^{\rm I}\approx \chi_{\rm ion}^{\rm
I}/\chi_{\rm e}^{\rm I}\approx \nu_{\rm ion}/\nu_{\rm e}$, so the
relative contribution of electron and ion to $\mathscr{D}^{\rm I}$
is determined by the relative importance of electron Landau damping
versus ion Landau damping. As shown in Fig.~\ref{Fig:ffSRS}, the
electron Landau damping reaches maximum when the phase velocity
$v_{\rm ph}$ is close to $v_{\rm the}$. Similarly, the ion damping
is maximum when the phase velocity $v_{\rm ph}$ is close to the ion
thermal velocity $v_{\rm th\alpha}$.
For SBS, typically $c_{\rm s}\ll v_{\rm the}$.
When $Z_\alpha$ increase, the ion acoustic velocity $c_{\rm s}$ also
increases, making the electron damping $\nu_{\rm e}$ increase
whereas the ion damping $\nu_{\rm ion}$ decrease when $v_{\rm th\alpha}<c_{\rm s}$.
The contribution of electron damping or ion damping can be quite different in low-Z and high-Z plasma.
Furthermore, the possible super-Gaussian exponent $m$ due to Langdon
effect is smaller in low-Z plasma than in high-Z plasma. Considering
the differences in low-Z plasma and high-Z plasma, in the following,
we discuss them separately.

\begin{figure}[!ht]
  \centering
  \includegraphics[angle=0,width=0.48\textwidth]{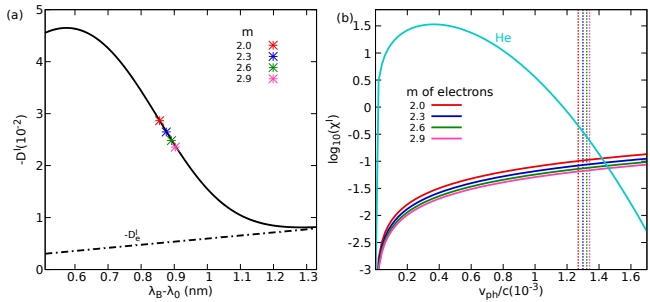}
  \caption{
    (a) $-\mathscr{D}^{\rm I}$ (solid line) and $-\mathscr{D}^{\rm I}_{\rm e}$ (dotted-dash line) versus wavelength shift $(\lambda_{\rm B}-\lambda_{\rm 0})$ for He plasma. The asterisk symbols indicate the wavelength location corresponding to the peak ${\rm K}_{\rm B}$ at $m=2$ (red), $m=2.3$ (blue), $m=2.6$ (green) and $m=2.9$ (magenta) in case `a'.
  (b) $\chi_\alpha^{\rm I} \propto -\partial f_{\alpha}^x/\partial v_x$ of electrons at different $m$ and He ions versus $v_{\rm ph}$. The IAW phase velocities corresponding to the peak wavelength for each $m$ in case `a' are indicated by the vertical dotted lines.
  }
  \label{Fig:ffHe}
\end{figure}
\begin{figure}[!ht]
  \centering
  \includegraphics[angle=0,width=0.48\textwidth]{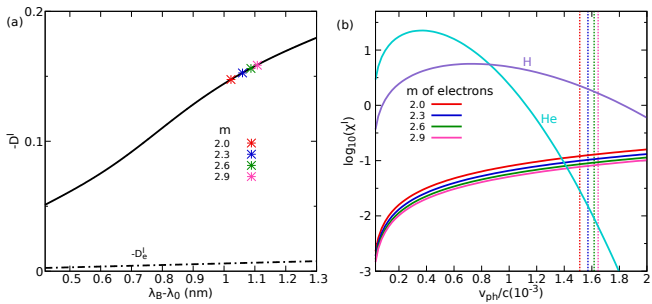}
  \caption{
    (a) $-\mathscr{D}^{\rm I}$ (solid line) and $-\mathscr{D}^{\rm I}_{\rm e}$ (dotted-dash line) versus wavelength shift $(\lambda_{\rm B}-\lambda_{\rm 0})$ for HHe (1:1) plasma. The asterisk symbols indicate the wavelength location corresponding to the peak ${\rm K}_{\rm B}$ at $m=2$ (red), $m=2.3$ (blue), $m=2.6$ (green) and $m=2.9$ (magenta) in case `a'.
    (b) $\chi_\alpha^{\rm I} \propto -\partial f_{\alpha}^x/\partial v_x$ of electrons at different $m$, H ions, and He ions versus $v_{\rm ph}$. The phase velocities of IAW corresponding to the peak wavelength for each $m$ in case `a' are indicated by the vertical dotted lines.
  }
  \label{Fig:ffHHe}
\end{figure}

In low-Z He and HHe plasmas, taking case `a' as example, we show
$-\mathscr{D}^{\rm I}$ in Fig.~\ref{Fig:ffHe}(a) and
Fig.~\ref{Fig:ffHHe}(a) for He and HHe plasmas, respectively. In
proximity to the peak wavelength, $-\mathscr{D}^{\rm I}$ is mainly
contributed by ions, because the ion damping which is proportional
to $\chi_{\rm ion}^{\rm I}$ is much higher than the electron damping
as shown in Figs.~\ref{Fig:ffHe}(b) and \ref{Fig:ffHHe}(b). Since
the Maxwellian distribution of ions is assumed, $\mathscr{D}_{\rm
ion}^{\rm I}$ itself does not depend on $m$. However, since the
phase velocity of IAW corresponding to the peak wavelength
$\lambda_{\rm B}$ increases with $m$, the decrease of
$-\mathscr{D}^{\rm I}_{\rm ion}$ with the peak wavelength
$\lambda_{\rm B}$ at different $m$ in He plasma leads to the
increase of the peak value of $\mathrm{K}_{\rm B}$ with $m$.
Similarly in HHe plasma, the increase of $-\mathscr{D}^{\rm I}_{\rm
ion}$ with the peak wavelength $\lambda_{\rm B}$ at different $m$
results in the decrease of the peak $\mathrm{K}_{\rm B}$ with $m$.

\begin{figure}[!ht]
  \centering
  \includegraphics[angle=0,width=0.48\textwidth]{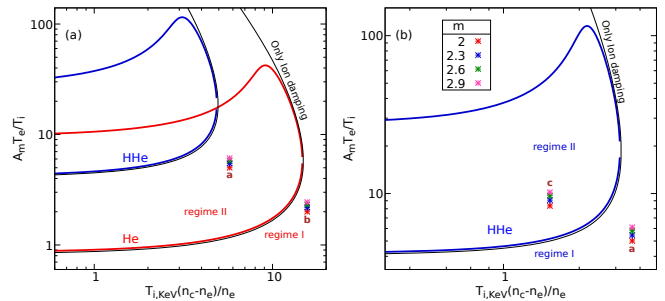}
  \caption{
(a) The boundary between regime I and II as specified by $A_mT_{\rm
e}/T_{\rm i}$ versus $T_{\rm i}(n_{\rm c}-n_{\rm e})/n_{\rm e}$ for
He (red line) and HHe (1:1) (blue line) plasmas.
(b) Zoomed in
boundary for HHe (1:1) (blue line) plasmas. The asterisk symbols
indicate different cases with conditions specified in
Table~\ref{Tab:caseHeHHe}, with the four points corresponding to
$m=2$, $m=2.3$, $m=2.6$ and $m=2.9$
from bottom to top for each case. The thin black curves show the
boundary with ion damping considered only.
  }
  \label{Fig:sepHHeI-II}
\end{figure}

It can be concluded that the variation trend of peak $\mathrm{K}_{\rm B}$ with $m$ is determined by
the behavior of $-\mathscr{D}^{\rm I}$ with $v_{\rm ph}$ near the phase velocity of IAW corresponding to the peak wavelength.
According to this, two parameter regimes can be defined.
In regime I, ${\partial \mathscr{D}^{\rm I}}/{\partial v_{\rm ph}}<0$ near the peak point and thus the peak $\mathrm{K}_{\rm B}$ decrease with $m$, while ${\partial \mathscr{D}^{\rm I}}/{\partial v_{\rm ph}}>0$ near the peak point and the peak $\mathrm{K}_{\rm B}$ increase with $m$ in regime II.
So, the boundary between regime I and II satisfies ${\partial \mathscr{D}^{\rm I}}/{\partial v_{\rm ph}}=0$ at the peak point
which maintains $\mathscr{D}^{\rm R}=0$.
\footnote{In fact, it can be proven under the condition ${\partial \mathscr{D}^{\rm I}}/{\partial v_{\rm ph}}=0$ and $\mathscr{D}^{\rm R}=0$, $\mathrm{K}_{\rm B}$ given in Eq.~(\ref{eq:KapEstimate}) satisfies $\partial \mathrm{K}_{\rm B}/\partial \omega_s=0$. That is, the peak condition of $\mathrm{K}_{\rm B}$ is exactly satisfied.
  So when the weak dependence of $-\mathscr{D}_{\rm e}^{\rm I}$ on $m$ can be neglected,
  these two conditions give the exact conditions for the boundary where the trend of the peak $\mathrm{K}_{\rm B}$ with $m$ changes.
}
These two equations
determine a relation between $A_mT_{\rm e}/T_{\rm i}$ and $T_{\rm
i}(1-n_{\rm e}/n_{\rm c})/n_{\rm e}$ for a given ion composition.
By numerical solution, the boundary between regime I and II for He
and HHe plasma are shown in Fig.~\ref{Fig:sepHHeI-II}, where
the weak dependence of $-\mathscr{D}_{\rm e}^{\rm I}$ on $m$ is ignored, and
the different cases shown in Fig.~\ref{Fig:SBSGainHeHHe} are also
labelled as asterisks.
As expected, case `b' is located in regime I of He plasma, and case
`c' is located in regime II of HHe plasma. Case `a' of He and HHe
are located in the special parameter range that is in regime II of
He plasma yet in regime I of HHe plasma, thus the variation
trend of peak $\mathrm{K}_{\rm B}$ with $m$ are opposite in case
`a' of He and HHe plasma even under the same parameter condition.

%
\begin{figure}[!h]
  \centering
  \includegraphics[angle=0,width=0.48\textwidth]{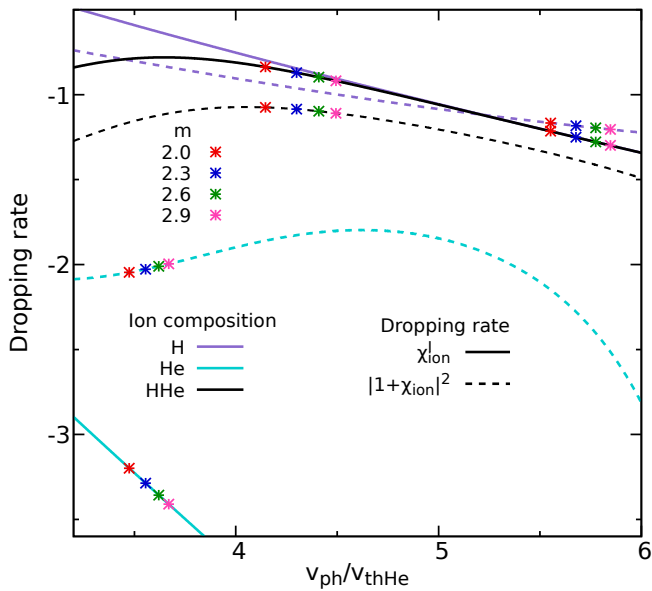}
  \caption{
    The dropping rate of $\chi_{\rm ion}^{\rm I}$ (solid line) and $|1+\chi_{\rm ion}|^2$ (dashed line)
    versus $v_{\rm ph}$ for fully ionized H (purple), He (blue) and HHe (1:1) (black) plasmas.
    The asterisk symbols indicate locations of the matched phase velocity corresponding to the peak of ${\rm K}_{\rm B}$ at $m=2$ (red), $m=2.3$ (blue), $m=2.6$ (green) and $m=2.9$ (magenta) in case `a'.
  }
  \label{Fig:XeIH_He_HHe}
\end{figure}

Now we take a closer look into the case `a' to comprehend the
physical mechanism for the effects of mixture.
Since $-\mathscr{D}_{\rm ion}^{\rm I}=\chi_{\rm ion}^{\rm I}/|1+\chi_{\rm ion}|^2$ dominates $-\mathscr{D^{\rm I}}$ near the matched $c_{\rm s}$ as discussed above, and both $\chi_{\rm ion}^{\rm I}$ and $|1+\chi_{\rm ion}|^2$ drop with $v_{\rm ph}$, the relative dropping rate of $\chi_{\rm ion}^{\rm I}$ and $|1+\chi_{\rm ion}|^2$ with $v_{\rm ph}$ determines the sign of ${\partial \mathscr{D}^{\rm I}}/{\partial v_{\rm ph}}$ and thus the regime.
In Fig.~\ref{Fig:XeIH_He_HHe}, the dropping rate of $\chi_{\rm ion}^{\rm I}$ and $|1+\chi_{\rm ion}|^2$, which are defined as $\partial (\ln\chi_{\rm ion}^{\rm I})/\partial v_{\rm ph}$ and $\partial (\ln|1+\chi_{\rm ion}|^2)/\partial v_{\rm ph}$ respectively, are shown as the function of $v_{\rm ph}$ for H, He and HHe (1:1) with the same plasma parameters in case `a', as well as the matched phase velocity corresponding to the peak ${\rm K}_{\rm B}$ at different $m$ for each ion composition. The $v_{\rm ph}$ and the dropping rate are normalized by thermal velocity of helium $v_{\rm thHe}$ and $1/v_{\rm thHe}$, respectively.
As expected, in HHe plasma, the matched $c_{\rm s,HHe}$ is between the matched $c_{\rm s,H}$ in single-species H plasma and $c_{\rm s,He}$ in single-species He plasma, with $c_{\rm s,HHe}/v_{\rm thHe}\sim 4$ and $c_{\rm s,HHe}/v_{\rm thH}\sim 2$.
Because $c_{\rm s,HHe}$ is closer to $v_{\rm thH}$, the Landau damping is mainly contributed by H ions, and the dropping rate of $\chi_{\rm HHe}^{\rm I}$ is closer to $\chi_{\rm H}^{\rm I}$ in case of single H ions, as shown in Fig.~\ref{Fig:XeIH_He_HHe}. However, due to the contributions of He ions, the dropping rate of $|1+\chi_{\rm HHe}|^2$ is larger than $|1+\chi_{\rm H}|^2$ of the single H case. As a consequence, the decrease of $\chi_{\rm HHe}^{\rm I}$ with $v_{\rm ph}$ is slower than the decrease of $|1+\chi_{\rm HHe}|^2$ in HHe plasma near $c_{\rm s,HHe}$, which leads to ${\partial \mathscr{D}^{\rm I}}/{\partial v_{\rm ph}}<0$ and thus in regime I in case `a'.

\begin{figure}[!ht]
  \centering
  \includegraphics[angle=0,width=0.48\textwidth]{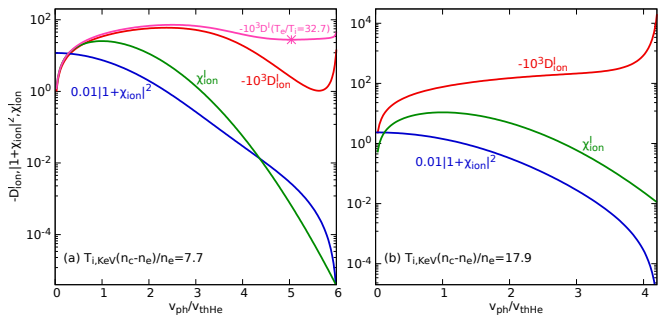}
  \caption{
    The typical variation of $-\mathscr{D}_{\rm ion}^{\rm I}$, $|1+\chi_{\rm ion}|^2$, and $\chi_{\rm ion}^{\rm I}$ with $v_{\rm ph}$ for He plasma with $T_{\rm i}(1-n_{\rm e}/n_{\rm c})/n_{\rm e}=7.7$ (a) and $17.9$ (b).
In the left panel, $\mathscr{D}^{\rm I}=\mathscr{D}_{\rm ion}^{\rm I}+\mathscr{D}_{\rm e}^{\rm I}$ with $T_{\rm e}/T_{\rm i}=32.7$ corresponding to the upper boundary of regime II at $T_{\rm i}(1-n_{\rm e}/n_{\rm c})/n_{\rm e}=7.7$ is also shown with its minimum point indicated by the asterisk symbol.
  }
  \label{Fig:DIonVar}
\end{figure}

%
In Fig.~\ref{Fig:sepHHeI-II}, the regime II is constrained between a
limited range of $A_{\rm m}T_{\rm e}/T_{\rm i}$ when $T_{\rm i}(1-n_{\rm e}/n_{\rm c})/n_{\rm e}$ is small, while it is totally
regime I when $T_{\rm i}(1-n_{\rm e}/n_{\rm c})/n_{\rm e}$ is large
enough. This general feature of the partition between regime I and
II can be explained as follows.
When $T_{\rm i}(1-n_{\rm e}/n_{\rm c})/n_{\rm e}$ is large enough, $-\mathscr{D}^{\rm I}_{\rm ion}$ increases with $v_{\rm ph}$ for all possible $v_{\rm ph}$ of IAW, as shown in Fig.~\ref{Fig:DIonVar}(b).
Since $-\mathscr{D}^{\rm I}_{\rm e}$ also increases (linearly) with $v_{\rm ph}$ for $v_{\rm ph}\ll v_{\rm the}$,
$-\mathscr{D}^{\rm I}$ increases with $v_{\rm ph}$  over the entire range of all possible $c_{\rm s}$,
rendering such $T_{\rm i}(1-n_{\rm e}/n_{\rm c})/n_{\rm e}$ belonging to regime I for all $A_mT_{\rm e}/T_{\rm i}$.
When $T_{\rm i}(1-n_{\rm e}/n_{\rm c})/n_{\rm e}$ is smaller, as $v_{\rm ph}$ increases, $-\mathscr{D}^{\rm I}_{\rm ion}$ increases to a maximum  firstly then decreases to a minimum, because $\chi_{\rm ion}^{\rm I}$ drops with $v_{\rm ph}$ faster and faster after $v_{\rm ph}>v_{\rm th\alpha}$,
and its dropping rate exceeds the dropping rate of $|1+\chi_{\rm ion}|^2$ beyond this maximum point, as shown in Fig.~\ref{Fig:DIonVar}(a). While the minimum point arises because $-\mathscr{D}^{\rm I}_{\rm ion}$ rises sharply towards a peak at large $v_{\rm ph}$  where $\chi_{\rm ion}+1$ approaches zero.
This peak corresponds to the greatest possible matched $c_{\rm s}$ when $A_mT_{\rm e}\gg T_{\rm i}$, and $\chi_{\rm e}\ll 1$ is negligible in the dispersion equation $\epsilon\approx 1+\chi_{\rm ion}=0$.
For a given $T_{\rm i}$, the matched $c_{\rm s}$ increases with the increased $A_{\rm m}T_{\rm e}/T_{\rm i}$, consequently, the ion damping decreases while the electron damping increases,
as shown in Figs.~\ref{Fig:ffHe}(b) and \ref{Fig:ffHHe}(b).
When $A_{\rm m}T_{\rm e}/T_{\rm i}$ is small enough, the contribution of $-\mathscr{D}^{\rm I}_{\rm ion}$ dominates.
The boundary coincides with the
boundary when only the ion damping considered alone as shown in Fig.~\ref{Fig:sepHHeI-II}.
The lower boundary corresponds to the maximum point of $-\mathscr{D}^{\rm I}_{\rm ion}$.
For a parameter point with $A_{\rm m}T_{\rm e}/T_{\rm i}$ located below the lower boundary, the matched $c_{\rm s}$ is on the left side of
the phase velocity corresponding to the maximum point of
$-\mathscr{D}_{\rm ion}^{\rm I}$, thus $-\mathscr{D}^{\rm I}$
increases with $v_{\rm ph}$ and this parameter point is in regime I.
While for a parameter point with $A_{\rm m}T_{\rm e}/T_{\rm i}$ located above the lower boundary, the matched $c_{\rm s}$ is on the
right side of the maximum point, thus $-\mathscr{D}^{\rm I}$
decreases with $v_{\rm ph}$ and this parameter point is in regime II.
With the increase of $A_{\rm m}T_{\rm e}/T_{\rm i}$, the matched $c_{\rm s}$ becomes closer and closer to the minimum point of $-\mathscr{D}_{\rm ion}^{\rm I}$, but $-\mathscr{D}_{\rm e}^{\rm I}$ which increases with $v_{\rm ph}$ becomes more and more important. At the upper right boundary of regime II with a relatively large $T_{\rm i}(1-n_{\rm e}/n_{\rm c})/n_{\rm e}$, $c_{\rm s}$ reaches the minimum point before $-\mathscr{D}_{\rm e}^{\rm I}$ becomes sufficiently important. While for smaller $T_{\rm i}(1-n_{\rm e}/n_{\rm c})/n_{\rm e}$,
the increase of $-\mathscr{D}_{\rm e}^{\rm I}$ with $v_{\rm ph}$ completely cancels the
decrease of $-\mathscr{D}_{\rm ion}^{\rm I}$ with $v_{\rm ph}$
before the phase velocity approaches the minimum point of $-\mathscr{D}_{\rm ion}^{\rm I}$ as shown in Fig.~\ref{Fig:DIonVar}(a), where the asterisk symbol indicates the location of the phase velocity corresponding to the minimum point of $-\mathscr{D}^{\rm I}$ with consideration of electron damping. This minimum point of $-\mathscr{D}^{\rm I}$ corresponds to the upper boundary of regime II at small $T_{\rm i}(1-n_{\rm e}/n_{\rm c})/n_{\rm e}$ in Fig.~\ref{Fig:sepHHeI-II}.

\begin{figure}[!ht]
  \centering
  \includegraphics[angle=0,width=0.48\textwidth]{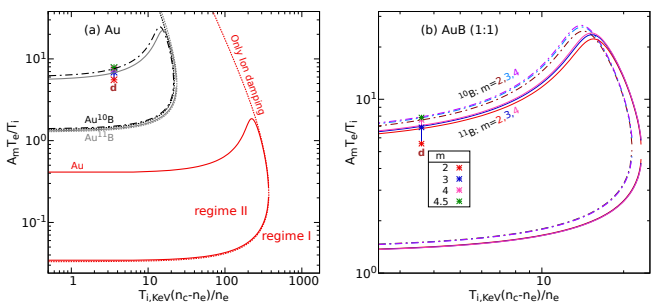}
  \caption{
(a) The boundary between regime I and II as specified by $A_mT_{\rm
e}/T_{\rm i}$ versus $T_{\rm i}(n_{\rm c}-n_{\rm e})/n_{\rm e}$ for
Au (red solid line), Au$^{10}$B (black dot-dashed line) and Au$^{11}$B
(gray solid line) plasma.
The dotted curves show the boundary with ion damping considered only.
(b) Zoomed in boundary for Au$^{10}$B (dot-dashed curves)  and
Au$^{11}$B (solid curves) plasma at different $m$.
The asterisk symbols indicate
the case `d' with conditions specified in Table~\ref{Tab:caseAuAuB},
with the four points corresponding to $m=2$, $m=3$,
$m=4$ and $m=4.5$ from bottom to top.
  }
  \label{Fig:sepAuBI-II}
\end{figure}

In high-Z plasmas, the Langdon effect can be more significant and
the super-Gaussian exponent can approach about $5$. With the same
definition of parameter regimes discussed above,
the boundary between regime I and II is shown in
Fig.~\ref{Fig:sepAuBI-II} for Au and AuB plasma (including both
Au$^{10}$B and Au$^{11}$B), respectively.
Since the weak dependence of $-\mathscr{D}_{\rm e}^{\rm I}$ on $m$, the parameter boundary shifts with $m$
indistinctly as shown in Fig.~\ref{Fig:sepAuBI-II}(b). Here, only the boundary at $m=2$ is shown for Au plasma.
It is noticed that when
the ion damping is considered alone, the regime II of AuB plasma is
enclosed in the regime II of Au plasma. In fact, this holds
generally for mixing low-Z species into the single ion species
plasma with higher charge state. However, taking into account
contributions from the electron damping, parameter region of regime
II of Au plasma is reduced, and the regime II of AuB plasma locates
inside the regime I of Au plasma. An interesting point of
Fig.~\ref{Fig:sepAuBI-II} is that the boundary of regime II for
Au$^{10}$B plasma is upwardly shifted
relative to that for Au$^{11}$B plasma. Such an isotope effect
should be embodied especially near the boundary of regime II. As
shown in Fig.~\ref{Fig:sepAuBI-II} case `d' belongs to regime I in
Au plasma, and spans both regime I and II in AuB plasma.
Thus, as expected, the peak $\mathrm{K}_{\rm B}$ decreases with $m$
monotonically in Au plasma, while the trend of the peak
$\mathrm{K}_{\rm B}$ with $m$ becomes non-monotonic and depends
sensitively on the isotopic type of the low-Z species $\rm B$ in AuB
plasma, as shown in Fig.~\ref{Fig:SBSGainAuAuB}.

\begin{figure}[!ht]
  \centering
  \includegraphics[angle=0,width=0.48\textwidth]{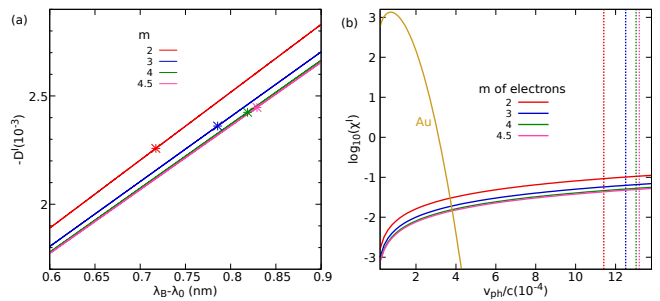}
  \caption{
    (a) $-\mathscr{D}^{\rm I}$ (solid line) versus wavelength shift $(\lambda_{\rm B}-\lambda_{\rm 0})$ for Au
    plasma at $m=2$ (red), $m=3$ (blue), $m=4$ (green) and $m=4.5$ (magenta).
    The asterisk symbols indicate the wavelength location corresponding to the peak ${\rm K}_{\rm B}$ in case `d'.
  (b) $\chi_\alpha^{\rm I} \propto -\partial f_{\alpha}^x/\partial v_x$ of electrons at different $m$ and Au ions.
  The IAW phase velocities corresponding to the peak wavelength for each $m$ in case `d' are indicated by the vertical dotted lines.
  }
  \label{Fig:ffAu}
\end{figure}

\begin{figure}[!ht]
  \centering
  \includegraphics[angle=0,width=0.48\textwidth]{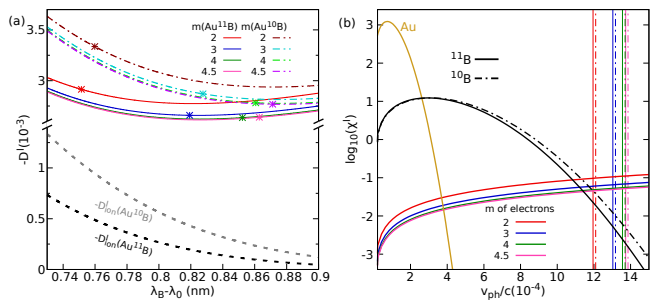}
  \caption{
    (a) $-\mathscr{D}^{\rm I}$ (solid and dash-dotted lines) and $-\mathscr{D}_{\rm ion}^{\rm I}$ (dashed line) versus wavelength shift $(\lambda_{\rm B}-\lambda_{\rm 0})$
    for Au$^{10}$B and Au$^{11}$B plasmas at different $m$.
    The asterisk symbols indicate the wavelength location corresponding to the peak ${\rm K}_{\rm B}$ at each $m$ in case `d'.
  (b) $\chi_\alpha^{\rm I} \propto -\partial f_{\alpha}^x/\partial v_x$ of electrons
  at different $m$, Au ions, $^{10}$B ions, and $^{11}$B ions.
  The IAW phase velocities corresponding to the peak wavelength for each $m$ in case `d' are indicated by the vertical dot-dashed lines for Au$^{10}$B plasma
and by the vertical solid lines for Au$^{11}$B plasma.
  }
  \label{Fig:ffAuB10B11}
\end{figure}

Using case `d' as example, we take a closer look into the effects of
the super-Gaussian EEDF and mixture near the upper boundary between
regime I and II. In Au plasma, the electron damping dominates over
the negligible ion damping since the matched $c_{\rm s}$ corresponding to
the peak ${\rm K}_{\rm B}$ is much larger than ion thermal velocity $v_{\rm thAu}$, $-\mathscr{D}^{\rm I}\approx -\mathscr{D}_{\rm e}^{\rm I}$ increases with the matched phase velocity $c_{\rm s}$ and
hence $\lambda_{\rm B}$, but decreases with $m$ as shown in
Fig.~\ref{Fig:ffAu}(b). The first effect has a greater impact in
case `d', which leads the peak ${\rm K}_{\rm B}$ decrease with $m$
as shown in Fig.~\ref{Fig:ffAu}(a). After mixing the low-Z boron
species into Au plasma, the variation of the ion damping with
$v_{\rm ph}$ due to B ions is not negligible compared to the
variation of electron damping with $v_{\rm ph}$, as shown in
Fig.~\ref{Fig:ffAuB10B11}(b). Furthermore, $-\mathscr{D}_{\rm ion}^{\rm I}$ decreases with $\lambda_{\rm B}$ whereas
$-\mathscr{D}_{\rm e}^{\rm I}$ increases with $\lambda_{\rm B}$. The
trend of the peak $\mathrm{K}_{\rm B}$ with $m$ depends on the
competition of the opposite change of $-\mathscr{D}_{\rm e}^{\rm I}$
and $-\mathscr{D}_{\rm ion}^{\rm I}$ with $\lambda_{\rm B}$.
In Au$^{11}$B plasma, this competition effect is the main reason for
the non-monotonic change of the peak ${\rm K}_{\rm B}$ with $m$.
Nevertheless, the obviously high $-\mathscr{D}^{\rm I}$ at $m=2$ as
shown in Fig.~\ref{Fig:ffAuB10B11}(a) leads the significant lower
value of peak ${\rm K}_{\rm B}$ at $m=2$ than $m=3,4,5$ as shown in
Fig.~\ref{Fig:SBSGainAuAuB}(b). Compared to $^{11}$B ion, the
lighter $^{10}$B ion has a larger thermal velocity as shown in
Fig.~\ref{Fig:ffAuB10B11}(b). As a result, near the peak wavelength,
the ion damping contributed by $^{10}$B is larger than $^{11}$B. On
one hand, this leads to a smaller ${\rm K}_{\rm B}$ in Au$^{10}$B
than in Au$^{11}$B plasma as shown in
Fig.~\ref{Fig:SBSGainAuAuB}(b). On the other hand, it makes the
contribution of $-\mathscr{D}_{\rm ion}^{\rm I}$ to
$-\mathscr{D}^{\rm I}$ become larger in Au$^{10}$B plasma, which
leads to a stronger decreasing trend of $-\mathscr{D}^{\rm I}$ with
$\lambda_{\rm B}$ as shown in Fig.~\ref{Fig:ffAuB10B11}(a).
Thus, the peak ${\rm K}_{\rm B}$ increase with $m$ in Au$^{10}$B
plasma as shown in Fig.~\ref{Fig:SBSGainAuAuB}(b).


%
\section{Discussion and summary}
\label{sec:conc} It is worthwhile to mention that in the
high-density gas-filled hohlraum experiments on NIF facility, the
peak wavelength of SRS is shorter than the simulated results
obtained by the inline ray-tracing model and artificial seed of SRS
backscattered light is needed to match the experimental reflectivity
~\cite{Strozzi2017InterPlayLPIHydro}. Commonly, the SRS of the inner
cone is mainly stimulated in the high density region with $n_{\rm
e}>0.1 n_{\rm c}$ and $T_{\rm e}<5~\rm KeV$ located in the small
$k_{\rm l}\lambda_{\rm De}$ regime where the wavelength of
backscattered light of SRS decreases with the super-Gaussian
exponent $m$. As a result, the simulated peak wavelength of SRS
light is expected to be shorter with the consideration of the
Langdon effect, which could be a possible explanation for this
discrepancy in SRS spectra. Besides, considering the enhancement of
SRS due to the Langdon effect in ray-tracing calculations may also
ameliorate the current problem of the underestimated SRS
reflectivity without the artificial seed. For SBS, the typical
parameters $T_{\rm e}/T_{\rm i} \sim 1.4\text{-}6$ and $T_{\rm
i}\sim \rm KeV$ in plasma ablated from hohlraum wall in experiments
are roughly located in regime II of the high-Z plasma mixed with
low-Z ions like AuB, but located in regime I of the single high-Z
plasma like Au. The Langdon effect itself can decrease the
convective growth of SBS in single high-Z plasma but enhance SBS in
mixed plasma, which may attenuate the improvement in suppression of
SBS by mixing low-Z ion species into the high-Z plasma. Here, the
influences of Langdon effect on SRS and SBS are mainly investigated
based on the linear convective gain obtained from Vlasov model. In
actual experiments, some possible nonlinear effects like electron
trapping and the three-dimensional distribution of the high
intensity speckles within the overlapped laser beams make the
coupling process of LPIs more complex, which still needs deeper
investigations in future.

In summary, the Langdon effect should be prevalent in ICF hohlraum
experiments and important to LPIs due to the sensitivity of LPIs on
EEDF.
Based on linear analysis, it is found that the peak wavelength of
the scattered wave and peak value of the spatial growth rate of both
SRS and SBS processes are observably influenced by the Langdon
effect which induces a super-Gaussian EEDF. For SRS, a
super-Gaussian EEDF modifies the dispersion relation of EPW,
yielding a redshift or blueshift of the peak wavelength. However,
the Landau damping of EPW is always reduced by the increased
super-Gaussian exponent $m$. Consequently, the peak spatial growth
rate of SRS always increases with $m$. For SBS, by reducing the low
energy electron number to shield the ions, a super-Gaussian EEDF
always increases the ion acoustic velocity, leading to a redshift of
SBS scattered wavelength. However, the effects of a super-Gaussian
EEDF on the peak spatial growth rate of SBS depend on the plasma
condition such as electron density, electron temperature and ion
temperature, and the ion composition (including the isotopic type)
in a complex way. The boundary between different regimes of the
enhancement or reduction of SBS by Langdon effect is given, and
distinct behaviors are presented for low-Z and high-Z plasma with
typical parameters in hohlraum plasma. The clarification of the
Langdon effect on SRS and SBS can make us better understand the
experimentally observed spectra and growth of SRS and SBS. Also it
provides valuable references for improvement of the physical
modeling and simulations of the LPI processes.


\section{acknowledgments}
This work was supported by the National Key R\&D Program of China
(Grant No.~2017YFA0403204), the Science Challenge Project (Grant
No.~TZ2016005), the National Natural Science Foundation of China
(Grant No.~11875093 and~11875091), and the Development Funds of CAEP
(Grant No.~CX20210040).

\appendix
\section{Proof that the ion acoustic velocity is monotonically increasing with $A_mT_{\rm e}$}
\label{app:csprove}
The ion acoustic velocity is given by Eq.~(\ref{eq:csTwo}) as $c_{\rm s}^2=(A\pm \sqrt{B})/2$,
with the plus for the fast mode, and the minus for the slow mode.
Firstly we define
\begin{equation}
  s_{\rm e}\equiv \frac{A_mT_{\rm e}}{\overline{Z}M_{\rm p}q}=\frac{A_mT_{\rm e}}{\overline{Z}M_{\rm p}(1+A_mk^2\lambda_{\rm De}^2)}\propto \frac{A_mT_{\rm e}}{1+\epsilon_0 k^2 A_mT_{\rm e}/n_{\rm e}e^2},
  \label{eq:defqT}
\end{equation}
It can be seen $s_{\rm e}$ is monotonically increasing with $A_mT_{\rm e}$.
On the other hand,
\begin{equation}
  \frac{\partial A}{\partial s_{\rm e}}=\overline{Z^2/A}>0
  \label{eq:PartDA}
\end{equation}
and
\begin{equation}
  \frac{\partial \sqrt{B}}{\partial s_{\rm e}}=\frac{1}{\sqrt{B}}
  \left[(\frac{\gamma_1}{A_1}-\frac{\gamma_2}{A_2})(\frac{f_1Z_1^2}{A_1}-\frac{f_2Z_2^2}{A_2})\frac{T_{\rm i}}{M_{\rm p}}+s_{\rm e}\left(\overline{Z^2/A}\right)^2\right]
  \label{eq:PartDB}
\end{equation}
Using Eqs.~(\ref{eq:defBCs}), (\ref{eq:PartDA}) and (\ref{eq:PartDB}), we obtain
\begin{equation}
  B(\frac{\partial A}{\partial s_{\rm e}})^2-(\sqrt{B}\frac{\partial \sqrt{B}}{\partial s_{\rm e}})^2=
  [2Z_1Z_2(\frac{\gamma_1}{A_1}-\frac{\gamma_2}{A_2})\frac{T_{\rm i}}{M_{\rm p}}]^2\frac{f_1f_2}{A_1A_2}\geq 0
  \label{eq:DAdiffB2}
\end{equation}
Therefore,
\begin{equation}
  \frac{\partial A}{\partial s_{\rm e}}\geq |\frac{\partial \sqrt{B}}{\partial s_{\rm e}}|
  \label{eq:DAdiffB}
\end{equation}
In other words,
\begin{equation}
  \frac{\partial c_{\rm s}^2}{\partial s_{\rm e}}= \frac{1}{2}[\frac{\partial A}{\partial s_{\rm e}} \pm \frac{\partial \sqrt{B}}{\partial s_{\rm e}}]\geq 0
  \label{eq:DAdiffB}
\end{equation}
So $c_{\rm s}$ is monotonically increasing with $s_{\rm e}$, which
is itself monotonically increasing with $A_mT_{\rm e}$.
\end{document}